\documentclass[10pt, journal]{IEEEtran}

\usepackage{cite}
\usepackage{amsmath,amssymb,amsfonts}
\usepackage{algorithmic}
\usepackage{graphicx}
\usepackage{chemformula}
\usepackage{textcomp}
\usepackage{xcolor}
\usepackage{lipsum}
\usepackage{float}
\usepackage{multirow}
\usepackage{booktabs}
\usepackage{epstopdf}
\graphicspath{{figures/}}
\DeclareGraphicsExtensions{.pdf,.jpg,.png}
\usepackage{algorithm,algorithmic}
\usepackage{soul}
\usepackage{color}
\usepackage{textcomp}
\usepackage{xspace}
\usepackage{mathtools}
\usepackage[caption=false, font=footnotesize]{subfig}
\DeclarePairedDelimiter{\ceil}{\lceil}{\rceil}

\def\BibTeX{{\rm B\kern-.05em{\sc i\kern-.025em b}\kern-.08em
    T\kern-.1667em\lower.7ex\hbox{E}\kern-.125emX}}
    
\begin{document}

\title{CRAFT: Criticality-Aware Fault-Tolerance Enhancement Techniques for Emerging Memories-Based Deep Neural Networks}
\author{\IEEEauthorblockN{Thai-Hoang Nguyen,~\IEEEmembership{Student Member,~IEEE,} Muhammad Imran, ~\IEEEmembership{Member,~IEEE,} Jaehyuk Choi,~\IEEEmembership{Member,~IEEE,} and
      Joon-Sung Yang,~\IEEEmembership{Senior Member,~IEEE}}%

\IEEEcompsocitemizethanks{\IEEEcompsocthanksitem T. H. Nguyen is with the Department of Electrical and Computer Engineering, Sungkyunkwan University, Suwon 16419, Korea.\protect
\IEEEcompsocthanksitem M. Imran is with the Department of Electrical Engineering, School of Electrical Engineering and Computer Science (SEECS), National University of Sciences and Technology (NUST), Islamabad 44000, Pakistan.\protect
\IEEEcompsocthanksitem J. Choi is with the Department of Semiconductor Systems Engineering, Sungkyunkwan University, Suwon 16419, Korea.\protect
\IEEEcompsocthanksitem J.-S. Yang is with the School of Electrical and Electronic Engineering and Department of System Semiconductor Engineering, Yonsei University, Seoul 03722, Korea (e-mail: js.yang@yonsei.ac.kr)}
\thanks{
This work was supported in part by the Institute of Information \& Communications Technology
Planning \& Evaluation (IITP) grant funded by the Korea Government (MSIT) under Grant 2022-0-00971;
in part by the Next Generation Intelligent Semiconductor Development by the Ministry of Trade,
Industry and Energy (MOTIE) under Grant 20011074; and in part by the Basic Science Research Program
through the National Research Foundation of Korea funded by the Ministry of Education under Grant
NRF-2020M3F3A2A01082326. The EDA Tools used in this work were supported by IDEC, Daejeon, South
Korea.}
}

\markboth{IEEE Transactions on Computer-Aided Design of Integrated Circuits and Systems}%
{Nguyen \MakeLowercase{\textit{et al.}}: CRAFT: Criticality-Aware Fault-Tolerance Enhancement Techniques for Emerging Memories-Base Deep Neural Networks}



\IEEEtitleabstractindextext{%
\begin{abstract} 
Deep Neural Networks (DNNs) have emerged as the most effective programming paradigm for computer vision and natural language processing applications. With the rapid development of DNNs, efficient hardware architectures for deploying DNN-based applications on edge devices have been extensively studied. Emerging Non-Volatile Memories (NVMs), with their better scalability, non-volatility and good read performance, are found to be promising candidates for deploying DNNs. However, despite the promise, emerging NVMs often suffer from reliability issues such as stuck-at faults, which decrease the chip yield/memory lifetime and severely impact the accuracy of DNNs. A stuck-at cell can be read but not reprogrammed, thus, stuck-at faults in NVMs may or may not result in errors depending on the data to be stored. By reducing the number of errors caused by stuck-at faults, the reliability of a DNN-based system can be enhanced. This paper proposes CRAFT, i.e., Criticality-Aware Fault-Tolerance Enhancement Techniques to enhance the reliability of NVM-based DNNs in the presence of stuck-at faults. A data block remapping technique is used to reduce the impact of stuck-at faults on DNNs accuracy. Additionally, by performing bit-level criticality analysis on various DNNs, the critical-bit positions in network parameters that can significantly impact the accuracy are identified. Based on this analysis, we propose an encoding method which effectively swaps the critical bit positions with that of non-critical bits when more errors (due to stuck-at faults) are present in the critical bits. Experiments of CRAFT architecture with various DNN models indicate that the robustness of a DNN against stuck-at faults can be enhanced by up to $10^{5}$ times on CIFAR-10 dataset and up to 29 times on ImageNet dataset with only a minimal amount of storage overhead i.e., 1.17\%. Being orthogonal, CRAFT can be integrated with existing fault-tolerance schemes to further enhance the robustness of DNNs against stuck-at faults in NVMs. 
\end{abstract}
}
\maketitle
\IEEEdisplaynontitleabstractindextext
\IEEEpeerreviewmaketitle
\vspace{1cm}
\IEEEraisesectionheading{\section{Introduction}\label{sec:introduction}}

\IEEEPARstart{D}{eep} Neural Networks (DNNs), a subset of Machine learning (ML) algorithms, have demonstrated impressive effectiveness in various applications such as computer vision, natural language processing, big data analysis and etc. A typical DNN consists of multiple hidden layers sandwiched between an input layer and an output layer. This hierarchical design allows DNN to solve complex programming tasks that appear to be infeasible with conventional programming approaches. However, despite the potential, DNNs often require enormous amount of computational power and hardware overhead, which makes it difficult to deploy them in real-time computing applications often running on mobile devices. As a result of the rapid development of DNNs, there is an enormous increase in the demand for efficient and scalable hardware architectures for DNNs' deployment. Realizing the high computational cost of DNNs, various methodologies have been proposed to achieve hardware-efficient architectures for DNNs \cite{han2015deep, 10.1007/978-3-319-46493-0_32}. These techniques often focus on reducing the storage required by DNNs through network compression \cite{han2015deep} and precision reduction \cite{10.1007/978-3-319-46493-0_32}. Such methods have proven to be efficient, however, DNNs often need to sacrifice accuracy in exchange for a reduced implementation cost in resource-constrained devices. 

\begin{figure*}[t]
    \centering
    \includegraphics[]{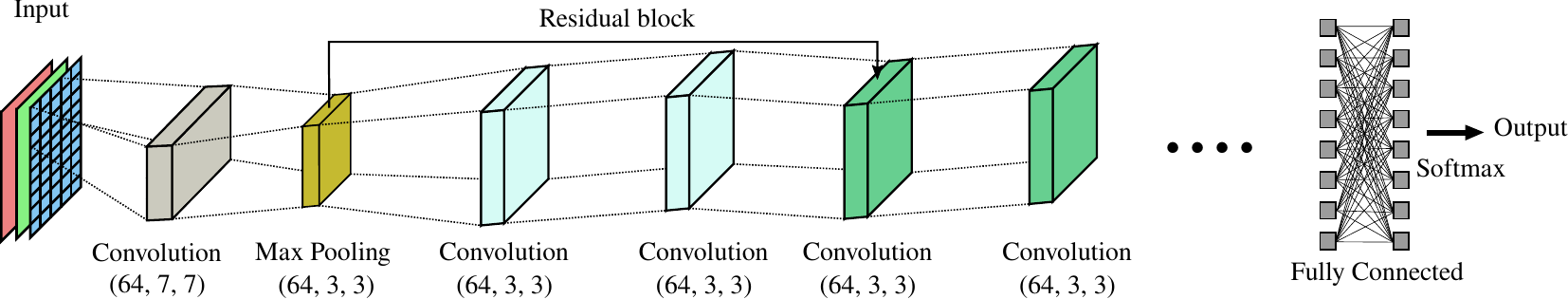}
    \caption{Typical Convolutional Neural Network architecture (ResNet-18\cite{he2016deep}) }
    \label{fig:resnet18-arch}
\end{figure*}

Memory plays a key role in the applications involving large amount of data like DNNs. Current
charge-based memory technologies such as Dynamic Random-Access Memory (DRAM), Static Random-Access
Memory (SRAM) and Flash are facing challenges in continuing technology scaling \cite{jeong2020pair}. Moreover, as the technology scales down, conventional memory technologies become highly prone to charge leakage which makes them a less attractive choice for data-intensive DNN applications. To cope with these issues posed by the conventional technologies, several emerging non-volatile memory technologies (NVMs) such as Resistive Random-Access Memory (ReRAM) and Phase Change Memory (PCM) have been extensively investigated over the past decade. With better scaling potential, better read performance and non-volatility \cite{7495087}, emerging NVMs are considered to be the potential replacement of the current charge-based memory technologies. 

Beside being used for storage, thanks to their analog characteristic, emerging NVMs have also played a major role in designing high-performance and energy-efficient In-memory Computing (ICM) based accelerators for DNNs \cite{7422838, shafiee2016isaac, joshi2020accurate}. Such accelerators use emerging NVMs cells (i.e., ReRAM, PCM) to store the network's parameters and perform the matrix-vector multiplication in-place by organizing NVMs cells in a crossbar manner. With in-place computations, NVMs based IMC architecture eliminates the data movement between memory and separate computing units, which is found to be very costly in conventional von Neumann architectures. These features make the emerging memories an ideal choice for the future hardware implementations of DNNs.

Despite their promising features, emerging NVMs often suffer from hard error (i.e., stuck-at faults)
\cite{freitas2008storage, ielmini2018memory, kwon_imran2021} due to their low endurance and immature manufacturing process. A stuck-at fault occurs when the resistance/conductance state of an emerging NVMs cell can not be changed by a write operation. A stuck-at cell can still be read but not reprogrammed, thus, errors caused by stuck-at faults only arise when the stuck-at cell's state is not aligned with the desired data. Building on this insight, several fault-tolerance techniques have been proposed to increase the lifetime of an emerging NVMs-based memory system  \cite{10.1145/1815961.1815980,5695530,6820868,6690215,10.1145/2897937.2897993}. In NVMs-based DNN architectures, despite the inherent fault-tolerance of DNNs, a small number of stuck-at NVMs cells (especially those corresponding to the critical bits) can still cause a catastrophic loss to DNN's accuracy \cite{mittal2020survey, li2017understanding}. Therefore, it is necessary to develop effective fault-tolerance enhancement techniques to mitigate such errors in NVMs-based DNN architectures. 

Existing works on tolerating stuck-at faults in emerging NVMs have aimed for neuromorphic applications  \cite{7926952,7858421,8060459,8119491,10.1145/3287624.3287707}. Despite being effective, these techniques often rely on an expensive retraining process of DNNs or a utilization of frequent auxiliary bits leading to a high storage overhead. On the other hand, several architectural techniques have also been proposed to tackle the problem of stuck-at errors in emerging NVMs, in general \cite{10.1145/1815961.1815980,5695530,6820868,6690215,10.1145/2897937.2897993}. Such techniques also require a large amount of hardware storage and complex encoding/decoding mechanisms, making them infeasible for resource-constrained hardware with real-time performance requirements. To address the problems of previous existing works, we propose multiple lightweight yet effective techniques, collectively named CRAFT, to tolerate errors caused by the stuck-at faults in NVMs based DNN architectures. The first technique, called Intra-Block Address Remapping, effectively remaps the weights inside a block of data so that the impact of stuck-at faults on DNN's accuracy is minimized. The second method addresses the problem of single-bit error by simply inverting the data in the data block. Results of these two techniques have been presented in our earlier work \cite{nguyen_imran2021}. To further enhance the robustness of the NVMs based DNNs, a novel Criticality-Aware Bits Switching method is proposed which further enhances the DNN's accuracy in the presence of stuck-at faults by addressing the bit criticality in DNNs. 

Rest of the paper is organized as follows. Section \ref{sec:background} covers the background of DNNs, emerging NVMs and stuck-at faults in emerging NVMs. Related works are presented in Section \ref{sec:related}. Section \ref{sec:proposed-techniques} introduces the proposed Criticality-Aware Fault-Tolerance Enhancement Techniques (CRAFT). Finally, we evaluate the effectiveness of the CRAFT against existing techniques in Section \ref{sec:evaluation}. Section \ref{sec:conclusion} concludes the paper. 

\section{Background}\label{sec:background}
\subsection{Deep Neural Networks (DNNs)}
Artificial neural networks (ANNs) are the computer algorithms inspired by the biological brain of animals. A layer of an ANN often consists of multiple nodes (i.e., neurons) connected to next layer through multiple connections (i.e., synapses/weights). Typical ANNs are made up of an input layer, an output layer and multiple hidden layers in between. A subset of ANN, Deep Neural Network (DNN), is an ANN with a large number of hidden layers (hence the name \textit{"Deep"} Neural Network). Over the last decade, DNNs have made major breakthroughs in various fields, rendering the conventional programming approaches in these domains as obsolete. Especially, in the field of computer vision, Convolutional Neural Networks (CNNs) \cite{he2016deep, han2017deep} have attracted a lot of interest due to their exceptional effectiveness.

Fig. \ref{fig:resnet18-arch} shows a typical CNN (ResNet-18) architecture. A CNN is often composed of three types of layers : Convolutional, Fully Connected (FC), Pooling and Normalization. The convolutional layer is often used for extracting features of the input data by convolving the input with multiple relatively small size filters. The output of the convolutional layer is then fed into the pooling layer to reduce the spatial size of the representation. In ResNet architecture, as shown in the figure, the output data is propagated through multiple residual blocks consisting of two convolutional layers and a shortcut connection. Such blocks allow the CNN to increase its depth (i.e., numbers of layer) while preventing the vanishing/exploding gradient effect. At the end of the network, data undergoes a fully connected layer for classification followed by a softmax layer which outputs the probability of each class.

The hierarchical structure of DNNs allows them to outperform the conventional programming algorithms in solving complex problems by breaking them into simpler ones. However, DNNs require immense hardware resources for storing parameters and performing computations. This makes it extremely challenging to deploy a large-scale DNN on resource-constrained hardware like mobile devices. To tackle this challenge, several techniques have been proposed to reduce the network size, thus making DNNs easier to deploy \cite{han2015deep}, \cite{krishnamoorthi2018quantizing}. The cost of using these techniques is the accuracy loss of DNNs. Depending on application, the accuracy loss may or may not be acceptable. In parallel with these approaches, several researches have investigated emerging non-volatile memories (NVMs) to provide high bandwidth, storage density and non-volatility for DNNs deployment. With such advantages over traditional charge-based memories, emerging NVMs are seen as ideal candidates for efficient and high-performance DNNs applications.

\subsection{Emerging Non-Volatile Memories (NVMs)}
Prominent emerging NVMs that are well-suited for DNN-based applications include Phase Change Memory
(PCM) and Resistive RAM (ReRAM) \cite{ielmini2018memory}. Phase Change Memory consists of a
chalcogenide phase-change material (e.g., \ch{Ge2Sb2Te5}) sandwiched between two electrodes. Data
can be stored in PCM by modulating the phase-change material's state which is either crystalline or
amorphous. The PCM cell has a low resistance in the crystalline state and a high resistance in
amorphous state. ReRAM, on the other hand, consists of a conducting material (typically \ch{HfO2})
placed in between two electrodes \cite{PAN20141}. The resistance state of a ReRAM cell can be
changed by altering the concentration of defects in the conductive filament. A high defects
concentration toward the bottom electrode changes the state of the ReRAM cell to the low-resistance
state and a high defects concentration towards the top electrode leads to the high-resistance state.
Both PCM and ReRAM have promising features of non-volatility, high switching speed and better
endurance compared to existing Flash memory \cite{ielmini2018memory}. However, due to certain
intrinsic characteristics of the underlying technology and an immature manufacturing process, these
memories often face reliability issues such as hard errors (stuck-at faults)
\cite{10.1145/1815961.1815980}, resistance drift \cite{8942113, , 8999530, thnguyen_2022, kwon2019pattern} and write disturbance
\cite{7926952, imran2021cent}. This poses a challenge when employing emerging NVMs for DNNs applications.

\begin{figure}[t]
	\centering
	\includegraphics[]{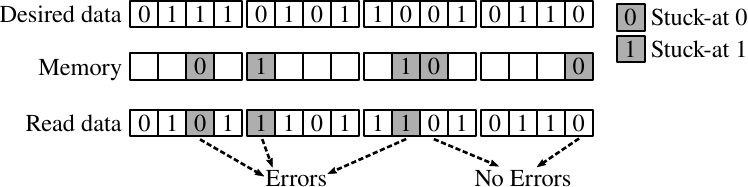}
	\caption{Example of Stuck-at faults in emerging NVMs}
	\label{fig:stuck-at-fault}
\end{figure}

\subsection{Stuck-at Faults and DNNs Accuracy}
\subsubsection{Stuck-at Faults in Emerging NVMs}\label{subsubsec:stuck-at-fault-in-nvms}
Stuck-at faults are a type of hard faults in emerging NVMs where the resistance state of a NVMs cell is locked at a certain state. When the cell resistance is fixed at the low resistance state, the fault is regarded as Stuck-at-Zero (SA0), otherwise, if the cell resistance is stuck at the high resistance state, this fault is considered to be Stuck-at-One (SA1). Depending on the data to be stored and the stuck-at state, a stuck-at fault may or may not cause error in the system. Fig. \ref{fig:stuck-at-fault} depicts a phenomenon of stuck-at faults in emerging NVMs. The first row shows the correct data expected to be stored and read from the memory. Second row shows the location and state of the stuck-at faults in emerging NVMs and the third row indicates the erroneous data read from the memory. As shown in the figure, the last two stuck-at fault locations do not introduce any error in the data because these cells' desired data is in-line with their stuck-at states. On the other hand, a mismatch between desired data and the stuck-at state causes error, which unintentionally flips the corresponding bit. Therefore, by aligning the desired data with the stuck-at resistance state, the number of readout errors in the system can be reduced. Previous works have used this property of stuck-at faults to mitigate their impact on the system \cite{6820868}.

According to the experiments using real fabricated ReRAM array in \cite{chen2014rram}, stuck-at zeros (SA0) and stuck-at ones (SA1) can be clustered in the entire column/row or distributed randomly in a ReRAM array. This stochastic nature of SAF makes it hard to model using any specific distributions, thus, many previous studies \cite{10.1145/3287624.3287707, 7926952} have chosen the uniform distribution to model SAFs to reduce complexity during the fault estimation process. The same consideration is also applied in our paper. Furthermore, since the proposed method does not focus on any specific type of eNVMs or SAFs, using uniform distribution for evaluations of SAFs allows CRAFT to be generalized and applicable for any use case. 

\subsubsection{Impact of Stuck-at Faults on Accuracy of DNNs}
\begin{figure}[t]
	\centering
	\includegraphics[]{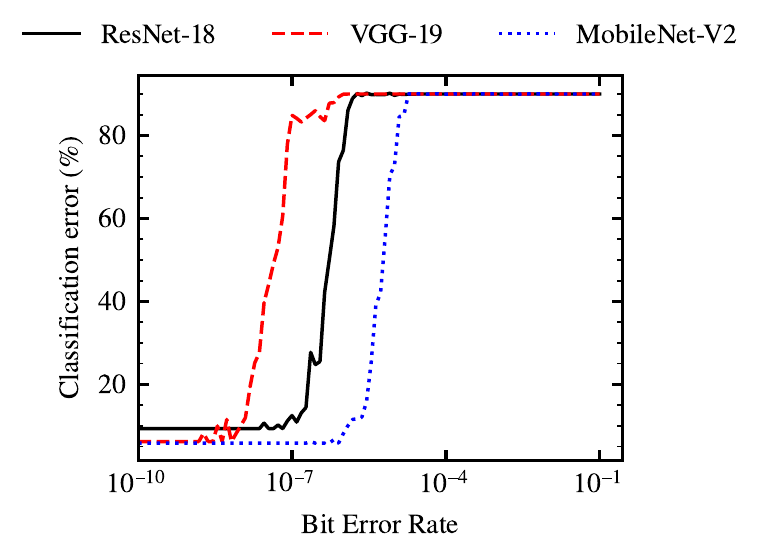}
	\caption{Impact of Stuck-at faults on accuracy of different DNN models (ResNet-18, VGG-19 and MobileNet-V2 on CIFAR-10) }
	\label{fig:stuck-at-fault-impact}
\end{figure}

A small number of stuck-at faults, especially in the critical bits, can cause a catastrophic change in DNN models accuracy \cite{li2017understanding, 10.1145/3287624.3287707}. Fig. \ref{fig:stuck-at-fault-impact} shows the impact of stuck-at faults on different DNN models' accuracy evaluated on CIFAR-10 (a popular image dataset for computer vision). As illustrated in the figure, when the Bit Error Rate (BER) for stuck-at faults increases to a certain point, the classification error of the DNN model increases exponentially. For example, for ResNet-18 (a state-of-the-art Convolutional Neural Network) \cite{he2016deep}, when the BER increases beyond $2\times10^{-6}$, the classification error stays at around 90\% level, which is the same as if the network is randomly guessing the results regardless of the input and trained parameters. Similar results are observed when considering different DNN models or datasets. Experimental details with additional results are discussed in Section \ref{sec:evaluation}.

\section{Related Works}\label{sec:related}
Several memory-centric works have been proposed to address the problem of stuck-at faults in emerging NVMs. Error Correcting Pointers (ECP) \cite{10.1145/1815961.1815980} detect and correct stuck-at faults by keeping the stuck-at cell address and data in additional storage. \cite{6820868} presents a method to enhance the correction capability of an ECC by a simple inversion operation. SAFER \cite{5695530} dynamically partitions the data such that only single error is presented in each partition and then uses a single-bit error correction code for recovery. \cite{qureshi2011pay} proposes a method to reduce the storage overhead of ECP by allocating different number of error correction entries to different lines according to the number of hard errors presented in the line. Although being effective in tolerating stuck-at faults of eNVMs-based system, these works often incur a large storage/hardware overhead, which is infeasible in the case of most resource-constrained edge devices. The proposed encoding techniques adds a minimum hardware overhead (1.17\% in terms of storage) to the system yet still efficiently enhances the robustness of DNN against SAFs.

Apart from memory-centric approaches, several works have been proposed to enhance the stuck-at faults tolerance capability of DNN based systems. The work in \cite{7926952} exploits the self-healing capability of DNNs and proposes a retraining method to reduce the impact of stuck-at cells in DNN accelerators. Such scheme can be effective in recovering the accuracy degradation from stuck-at errors; however, re-training is required when implementing these techniques, which is difficult to do when the DNN has been deployed to the edge devices. \cite{liu2019fault} redesigns the traditional error correction output code of DNNs using a collaborative logistic classifier, thus enhancing the DNN robustness against stuck-at faults. Despite being effective, this work also requires re-training (i.e., fine-tuning) to recover the accuracy impacted by SAFs. The need for re-training does not apply to the proposed fault-tolerant technique in this paper, since it is designed specifically for DNN inference on resource-constrained edge devices.

More relevant to our proposed techniques, many data remapping and redundancy based techniques have
been proposed \cite{7858421, 8060459, 8119491}. Specifically, \cite{7858421} introduces
mapping technique and redundant crossbar arrays to compensate the accuracy loss of DNN model caused
by the stuck-at faults in ReRAM crossbar array. Since this technique utilizes redundant crossbar
array that has the same size as the original array, the storage overhead and energy consumption of
such technique is considerably large. \cite{8060459} classifies weights according to their
criticality to the model's accuracy, remaps the significant weights to the fault-free memory cells
and fine-tunes the DNN model to enhance the accuracy. By relying on the criticality of DNNs to
address SAFs, such work is able to ease the re-training process and reduce storage overhead.
Nonetheless, the storage overhead caused by such technique can still be as large as 5\%, which is
much higher compared to our proposed technique. The method in \cite{10.1145/3287624.3287707} uses
matrix transformations to make the weights in the ReRAM crossbar array more robust to stuck-at
faults. Similar to other redundancy-based methods, \cite{10.1145/3287624.3287707} also
adds a significant amount of hardware overhead compared to the proposed technique. Specifically,
such a scheme can come at the expense of 8.19 $\times$ power consumption and 9.23 $\times$ area
overhead. The proposed technique in this paper only adds six additional bits for
encoding/decoding, making it highly efficient in term of storage/energy overhead

The existing memory-centric methods as well as DNN focused techniques often either require a large amount of additional hardware overhead or costly retraining process. In this paper, we propose a set of techniques that incur only minimal hardware overhead yet effectively enhance the fault-tolerance capability of DNNs in the presences of stuck-at faults. Moreover, the proposed techniques are orthogonal to the existing methods and can be implemented together to further enhance the robustness of DNNs.

\begin{figure}[t]
	\centering
	\includegraphics[]{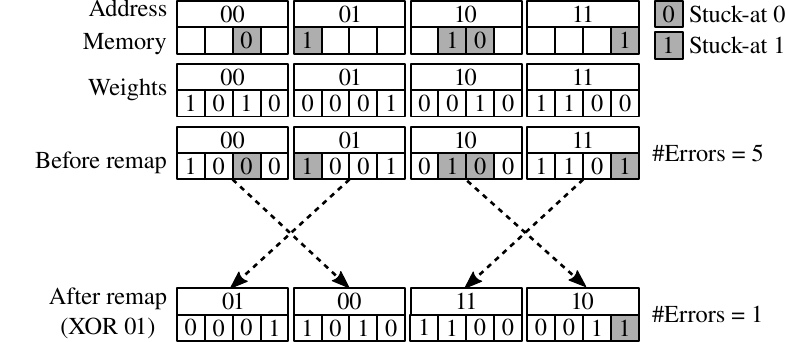}
	\caption{Example of Intra-Block Address Remapping technique}
	\label{fig:remapping-example}
\end{figure}

\section{CRAFT: Criticality-Aware Fault-Tolerance Enhancement Techniques}\label{sec:proposed-techniques}
The state of a stuck-at cell can be detected (by a read operation) but cannot be re-programmed to a different state. Leveraging this fact, we present multiple remapping and encoding techniques, collectively named CRAFT, to reduce the number of stuck-at errors in the DNN parameters (weights and biases). The proposed fault-tolerance techniques include \textit{Intra-Block Address Remapping}, \textit{Weight Inversion} and \textit{Criticality-Aware Bits Switching}.

\begin{figure*}[t]
	\centering
	\includegraphics[]{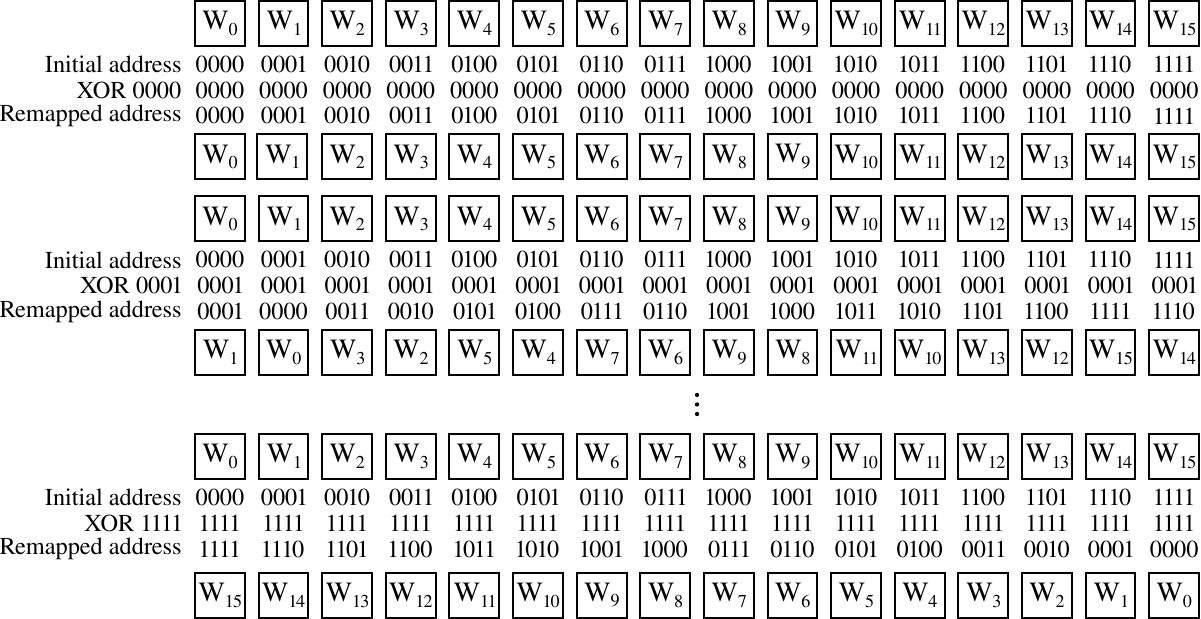}
	\caption{Intra-Block Address Remapping for 16 weights using 4-bit XOR operation}
	\label{fig:remapping-4-bit}
\end{figure*}

\subsection{Intra-Block Address Remapping}\label{subsec:weight-remapping}
The parameters (weights and biases) of DNN are often stored as a group in a data block. For instance, a typical 64B cache line sized data block can store up to sixteen 32-bit floating point weights/biases. The proposed remapping method operates within the typical data block to preserve memory access locality. Fig. \ref{fig:remapping-example} shows a simple example of 2-bit address remapping using the proposed \textit{Intra-Block Address Remapping} technique. An XOR operation of the address of each weight within a data block remaps the weight to a different location within the same block. This remapping allows to reduce the stuck-at faults by increasing the number of stuck-at cell states aligned with the desired bits. As shown in the figure, without remapping, the number of error due to stuck-at cells is five. After the proposed remapping technique (XOR the address with 01) is applied, the resulting errors due to stuck-at faults is reduced to one only. By considering multiple mappings through different XOR operations, the number of errors can be further reduced. The proposed method finally chooses the mapping which causes minimal impact (instead of the fewest errors, as explained in the next section) on DNNs accuracy. Fig. \ref{fig:remapping-4-bit} illustrates the proposed remapping technique using 4-bit XOR operation. As shown in the figure, sixteen different mappings can be obtained using 4-bit XOR operation.

\begin{figure}[t]
	\centering
	\includegraphics[]{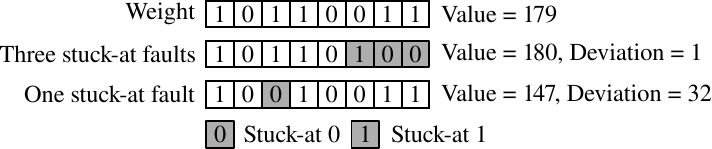}
	\caption{Example of fault significance in a data block. Three errors in insignificant bit positions result in insignificant change in value while one error in significant bit position causes large deviation from the actual value.}
	\label{fig:fault-significance}
\end{figure}

\subsubsection{Minimizing the Impact of Faults on DNN's Accuracy}\label{sec:minimize-impact} Errors that occur in significant bits of DNN parameters are more harmful to the network accuracy than the errors in insignificant bits. This can be easily understood using a simple example shown in Fig. \ref{fig:fault-significance}. As illustrated in the figure, frequent stuck-at faults in the insignificant bit positions result in only a small change from the actual weight value while fewer faults in the significant bit positions cause a greater change to the weight value. This is indicated by measuring the deviation in decimal value of the erroneous weight from that of the actual weight. As shown in Fig. \ref{fig:fault-significance}, three stuck-at faults in the insignificant bit position result in deviation of 1 (in decimal), while a single stuck-at fault in the significant bit position causes a deviation of 32 (in decimal). In floating-point representation, this criticality difference is even more evident due to the greater difference in the significance of exponent bits as compared to the other bit positions. Furthermore, in case of DNN, certain weights are more critical than the other weights, thus, merely minimizing the number of faults in the memory would not be always helpful. Therefore, the proposed remapping method chooses to minimize the deviation (from the original value) of weights instead of simply minimizing the number of stuck-at faults presented in memory. The proposed deviation minimization technique can be formulated as:

\begin{equation}
\label{remap_fomular}
w' \leftarrow w_{ri} \; \text{s.t} \; \delta = \min_{\delta \in \Delta} \sum_{i=0}^{N}|w_{ri}-w_{oi}|
\end{equation}
where $w'$ is the final weight that is used for inference, $w_{ri}$ and $w_{oi}$ refer to the new weight after remapping (obtained while considering stuck-at faults) and the original weight, respectively. $N$ is the number of weights within a selected data block. $\delta$ is the minimum net deviation and $\Delta$ is the set of all possible net deviations for different remappings of the weights. 

\begin{figure}[t]
	\centering
	\includegraphics[]{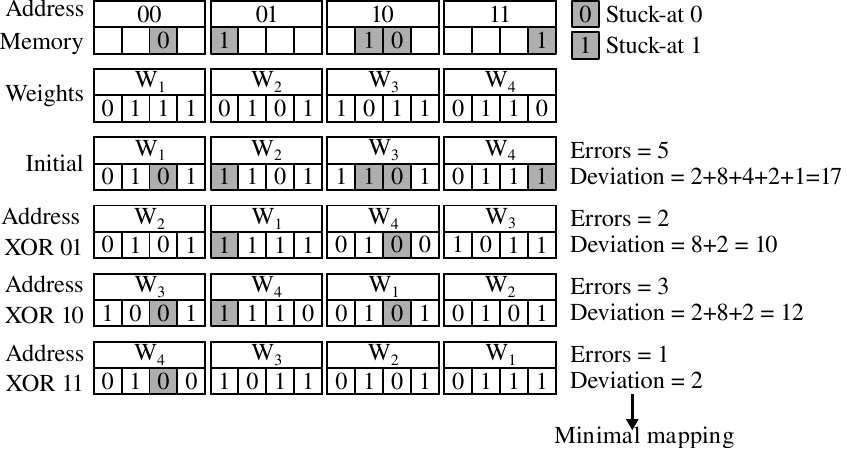}
    \caption{An example of \textit{Intra-Block Address Remapping} technique with deviation minimization. Net deviation for each mapping is calculated by summing up all weight value deviations in the data block. Mapping with minimum net deviation is chosen as minimal mapping for inference.}
	\label{fig:encoding-example}
\end{figure}

The proposed \textit{Intra-Block Address Remapping} technique with a deviation minimization algorithm is illustrated in Fig. \ref{fig:encoding-example}. For illustration simplicity, four 4-bit weights with 2-bit XOR operation for address remapping are depicted in the example. Five stuck-at faults (three SA1 and two SA0) are randomly distributed across the memory block, as shown in the figure. Initially, without any remapping, the readout data results in five errors which leads to the net deviation of 13 (in decimal). Using 2-bit XOR operation, four possible mappings can be obtained. As illustrated in the figure, the mapping which uses XOR 11 operation leads to the minimum net deviation (= 2 in decimal) from the actual weight. Therefore, the proposed technique chooses this mapping as the minimal mapping to store weights in the stuck-at-faults-prone emerging NVMs. 

\begin{figure}[t]
	\centering
	\includegraphics[]{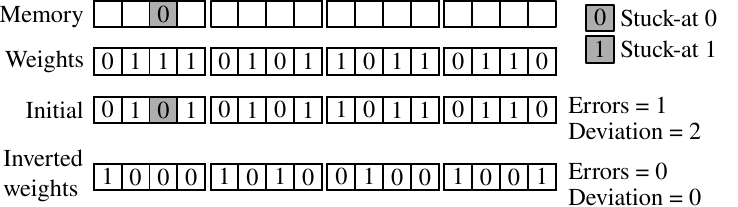}
	\caption{An example of the proposed Weight Inversion encoding technique}
	\label{fig:weight-inversion-example}
\end{figure}

\subsection{Weight Inversion}\label{subsec:weight-inversion}

When a data block contains only a single stuck-at fault, the error caused by this fault can be tolerated by simply inverting the data block. Based on this intuition, to further enhance the error-tolerance of a DNN architecture, a simple weight inversion encoding is proposed. Fig. \ref{fig:weight-inversion-example} illustrates an example of the proposed \textit{Weight Inversion} technique. As shown in the figure, an inversion operation is incorporated when there is a single stuck-at fault in the data block. After inversion and decoding, the read data results in zero error and zero deviation from the actual weights. The \textit{Weight Inversion} method is combined with the \textit{Intra-Block Address Remapping} technique by considering the possible mappings with original weight values as well as with the inverted weight values. 

\subsection{Criticality-Aware Bits Switching} \label{subsec:weight-rotation}
The proposed \textit{Intra-Block Address Remapping} and \textit{Weight Inversion} techniques efficiently enhance the fault-tolerance of DNNs by orders of magnitude, as shown by the evaluation results in Sec. \ref{sec:evaluation}. However, by only minimizing the net deviation between the erroneous and the actual weight values, errors that are critical to DNNs and unmaskable using these techniques can still significantly impact the DNNs accuracy. To address this, we introduce another encoding technique, incorporated on top of the first two techniques, that focuses on minimizing errors in the critical bit positions of the DNNs. The bit position criticality of DNNs against stuck-at faults is first analysed. Based on the results of this analysis, the \textit{Criticality-Aware Bits Switching} technique is proposed.

\subsubsection{Bit Position Criticality in DNNs}\label{subsubsec:bit-pos-criticality}

\begin{figure}[t]
	\centering
	\subfloat[]{\includegraphics{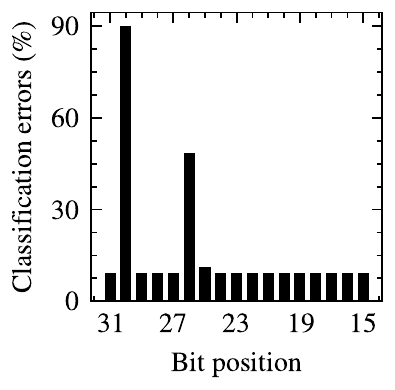}\label{fig:bitsens-32b}}
	\qquad
	\subfloat[]{\includegraphics{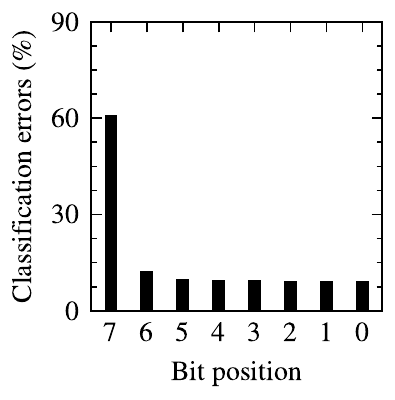}\label{fig:bitsens-8b}}
	\qquad
	\caption{Bit position criticality of DNNs against stuck-at faults. Each experiment is performed on ResNet-18 using CIFAR-10 dataset. The classification error of each bit position is averaged over 100 iteration.  (a) 32-bit Floating Point network  (b) 8-bit Quantized network}
	\label{fig:bit-sens}
\end{figure}

Certain bit positions in DNNs' parameters are more significant to the accuracy than the rest of bit
positions. In general, as discussed in Sec. \ref{sec:minimize-impact}, errors in the higher order
bit positions (MSBs) cause a greater impact to DNNs accuracy than errors in lower orders bit
positions (LSBs). However, in order to design an efficient error-tolerance mechanism for DNNs,
quantification of bit position criticality is also necessary. Fig. \ref{fig:bit-sens} shows an
experiment in which each bit position of DNNs' parameters is randomly disturbed with stuck-at
faults. The experiments are performed on ResNet-18 using CIFAR-10 dataset with the classification
error corresponding to each bit position averaged over 100 iterations. DNN parameters in both
floating-point and quantized precisions are considered. Fig. \subref{fig:bitsens-32b} shows the bit
position criticality of the full-precision (32-bit) network and Fig. \subref{fig:bitsens-8b}
illustrates the bit sensitivity of unsigned 8-bit quantized network. The stuck-at bit error rate
(BER) is fixed (arbitrarily) as $10^{-3}$ for both networks. 

As seen in the figures, only a few MSBs errors can cause a high impact to DNNs' accuracy while
errors in the other bit positions have a negligible impact on the accuracy. Specifically, for
network with 32-bit floating point parameters (Fig. \subref{fig:bitsens-32b}), two MSBs positions
that cause significant impact to DNN's accuracy are the 30\textsuperscript{th} and
26\textsuperscript{th} bit position. The reason for that can be explained by considering the DNN's
weight distribution. As shown in previous work \cite{8806855}, all weights in the DNN have a value
less than 1, hence, the value of the 30\textsuperscript{th} bit is always fixed at 0. This explains
why a bit flip error in the 30\textsuperscript{th} bit causes a catastrophic accuracy degradation.
Another bit position that can cause a severe impact to DNN's accuracy would be the second-zero
occurrence bit (SZOB) position (considering that the 30th bit is the first-zero occurrence
position). As reported in \cite{8806855}, the SZOB can be in the 25\textsuperscript{th},
26\textsuperscript{th} or 27\textsuperscript{th} bit position depending on the network, e.g., the
SZOB is found at 26\textsuperscript{th} bit position in 96\% of the weights in AlexNet and VGG16. As
illustrated in Fig. \ref{fig:bit-sens}, the SZOB of ResNet-18 is found to be at the
26\textsuperscript{th} position, and thus causes a severe impact to DNN's accuracy compared to other
bit positions. It is also worth noting that bit flip errors occurring in the sign bit (bit
31\textsuperscript{st}) show an insignificant influence on the accuracy. This is because most
weights in DNN have small value and thus would cause a small change in deviation when the sign is
flipped. For e.g., if a weight value is equal to 0.05 and a sign bit error makes the weight changed
to -0.05, the deviation in term of absolute value is only 0.1, which can be tolerable by the DNN.
The same case can happen when errors occur in the mantissa bits (bit
0\textsuperscript{th}--24\textsuperscript{th}) which can only produce a maximum deviation of 1.5. On
the other hand, if an error occurs in the exponent bit of the weight, the deviation between error
weight and original weight can be as large as $3.40\times10^{38}$ which can severely impact the
accuracy of DNN, as in case of bit 30\textsuperscript{th} shown in Fig. \subref{fig:bitsens-32b}.

For 8-bit quantized network, because unsigned 8-bit representation is used in this experiment, it is seen that only the 7\textsuperscript{th} bit position (MSB) causes a severe degradation to the network accuracy. Note that 32-bit floating point network's accuracy is deteriorated much more than that of 8-bit quantized network. This is due to the difference in the range of representation for each parameter in the network. Parameters in 8-bit quantized networks have a much smaller dynamic range than those of the full-precision network and thus cause a higher drop in the network accuracy. Similar observation has also been made in several previous works such as \cite{li2017understanding, mittal2020survey} as well as our evaluation result in Sec. \ref{sec:evaluation}. The results shown in Fig. \ref{fig:bit-sens} are found to be consistent for different DNN models and datasets, therefore, the aforementioned observation can be applicable for any configuration of DNNs. Based on the analysis of bit criticality, in the next section, we propose a novel \textit{Criticality-Aware Bits Switching} technique for enhancing the stuck-at fault-tolerance in DNNs.

\subsubsection{Criticality-Aware Bits Switching}

\begin{figure}[t]
    \centering
    \includegraphics[]{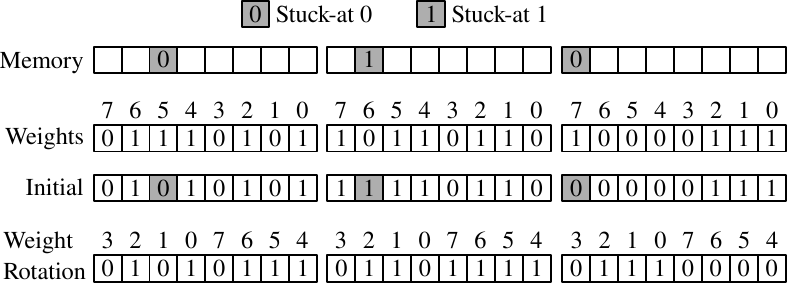}
    \caption{Example of \textit{Criticality-Aware Bits Switching}. MSBs positions are switched with LSBs positions so that the impact of errors in MSBs is minimized.}
    \label{fig:rotation-encoding}
\end{figure}

\begin{figure*}[t]
	\centering
	\subfloat[]{\includegraphics{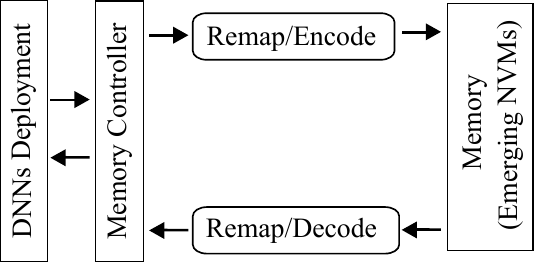}
	\label{fig:overall-architecture}}
	\qquad
	\medskip
	\subfloat[\label{fig:logic}]{\includegraphics{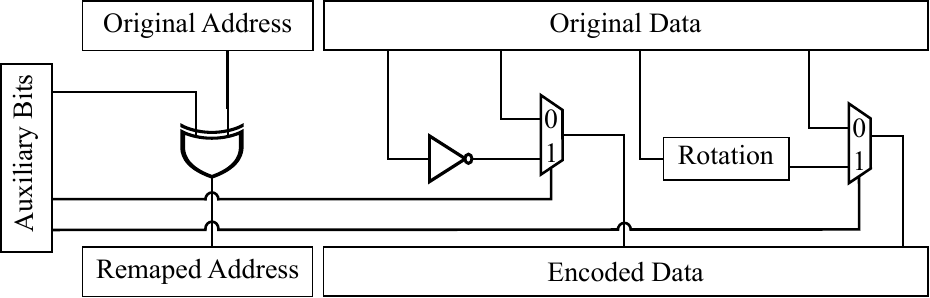}}
	\caption{Implementation of CRAFT  (a)  Overall Architecture  (b) Remapping and Encoding Logic}
	\label{fig:implementation}
\end{figure*}

Since the previously introduced techniques address stuck-at fault on a data block level, it is not guaranteed that errors in critical bits of each weight will be tolerated. Therefore, the proposed \textit{Criticality-Aware Bits Switching} technique operates on each individual weight in order to minimize the number of critical stuck-at errors within each weight. Since applying bits switching to each individual weight would require more auxiliary bits to encode/decode the data, the proposed method uses one unified switching operation for all weights within a data block to keep the storage overhead minimal. By doing this, the method only requires one bit per (512-bit) block for encoding/decoding which is a negligible hardware overhead. Nevertheless, the proposed method is flexible and can be applied in a more fine-grained manner with additional auxiliary bits to achieve greater robustness. 

\begin{small}
\begin{table*}[t]
	\centering
	\caption{Specifications of Evaluated DNNs and Datasets}
	\begin{tabular}{cccccc} 
		\toprule
		Dataset & Model    & Input Dimension & Parameters size (MB) & Classification Error ($\%$) & Precision\\ 
		\midrule
		\multirow{3}{4.5em}{CIFAR-10} & ResNet-18 & 3x32x32         & 43           & 9.31                  & 32-bit floating-point \\ 
		& VGG-19    & 3x32x32         & 533          & 6.19                  & 32-bit floating-point \\
		& MobileNet-V2 & 3x32x32      & 9            & 5.83                  & 32-bit floating-point \\
		\midrule
		\multirow{3}{4em}{\centering Imagenet} & ResNet-50 & 3x224x224       & 25.6          & 24.02                  & 8-bit quantized \\
		& VGG-19    & 3x224x224       & 143.72        & 27.64                  & 8-bit quantized\\ 
		
		& Inception-V4    & 3x224x224       & 43.03        & 20.06                  & 8-bit quantized\\
		\bottomrule
	\end{tabular}
	
	\label{table:1}
\end{table*}
\end{small}

Fig. \ref{fig:rotation-encoding} depicts an example of the proposed \textit{Criticality-Aware Bits Switching} technique. For the sake of illustration, three 8-bit weights with three critical bit positions being stuck-at faults are shown. The proposed method is incorporated by rotating the four MSBs (7\textsuperscript{th}, 6\textsuperscript{th}, 5\textsuperscript{th}, 4\textsuperscript{th}-bit) of each weight to the left and four LSBs (3\textsuperscript{rd}, 2\textsuperscript{nd}, 1\textsuperscript{st}, 0\textsuperscript{th}-bit) to the right. For example, in the first weight, the original data is "0111 0101" (in binary). When programmed to emerging NVMs, a stuck-at cell in the 5\textsuperscript{th} bit position will cause a net deviation of $2^{5}=32$ (in decimal) which in turns causes a severe impact on DNNs accuracy. After criticality-aware bit switching, the encoded data becomes "0101 0111" and the error caused by stuck-at fault in the critical cell is eliminated when reading from the data block. Intuitively, by switching MSBs with LSBs, errors in the MSBs and LSBs are also switched and thus, the method reduces the impact of errors in the MSBs significantly. It is important to note that, by doing intra-weight rotation, errors in the LSBs can also be increased. However, as discussed in the previous section, these errors does not cause a high impact on DNN's accuracy and the overall deviation in weight value would still be smaller. Based on the observation in the previous section, for the networks that uses 8-bit quantized parameters, four MSBs are switched with four LSBs. For 32-bit floating point networks, any MSBs rotation greater than five bits can be effective for MSBs fault-tolerance. In the proposed method, we choose to rotate ten MSBs with the LSBs in 32-bit weights to provide enough safety of margin. By combining the proposed \textit{Criticality-Aware Bits Switching} technique with the net deviation minimization, the robustness of the DNNs can be enhanced significantly, as shown in the evaluation results in Sec. \ref{sec:evaluation}.

\subsection{Implementation and Overhead}\label{subsec:implementation-overhead}
Since most of DNN applications are often trained once and then deployed to multiple edge devices, the emerging NVMs can be programmed after the training process is done. This eliminates a deployment-time overhead of the proposed system. When used only for inference, the proposed techniques do not have any significant impact on performance because the remapping and encoding operations add only a few logic gate delays to the critical path. 

Fig. \ref{fig:implementation} illustrates the implementation of the proposed CRAFT architecture. Specifically, Fig. \subref{fig:overall-architecture} and Fig. \subref{fig:logic} show the overall architecture and the remapping+encoding logic of CRAFT, respectively. During inference, the DNNs parameters read from the emerging NVMs get remapped/decoded based on the mapping/encoding that leads to minimal impact of stuck-at faults. As shown in Fig. \subref{fig:logic}, the \textit{Intra-Block Address Remapping} technique uses simple XOR logic to map the address to the new address while the \textit{Weight Inversion} method uses NOT operation to flip the data. The \textit{Criticality-Aware Bits Switching} technique uniformly rotates the weight data by simple rewiring logic. These remapping and encoding operations being trivial add negligible timing overhead during inference. 

The exact storage overhead of CRAFT can be calculated by dividing the number of auxiliary bits with the total number of data bits. For a typical 64B data block size and 32-bit floating-point DNN weights, CRAFT adds four auxiliary bits for address remapping, one bit for weight inversion and one bit for weight rotation (bits switching). Thus, the proposed techniques in CRAFT incur only 1.17\% storage overhead which is negligible. The effectiveness of CRAFT can be further improved with more fine-grained implementation while using smaller data blocks for remapping/encoding at the cost of increase in the number of auxiliary bits.

\section{Evaluation}\label{sec:evaluation}
The proposed stuck-at faults tolerance techniques are evaluated using different experiments that consider various DNN models, datasets and parameter configurations. In the following, details of the experimental setup followed by the evaluation results and discussion are presented. 

\subsection{Experimental Setup}

The simulations for stuck-at faults in DNNs are performed by using the Pytoch framework \cite{paszke2019pytorch}. All simulations are performed on an Intel\textsuperscript\textregistered \enspace Xeon\textsuperscript\textregistered \enspace CPU E5-1650 v4 with two Nvidia Titan XP GPUs having 24Gb of RAM. Table \ref{table:1} lists the specifications of the evaluated DNN models and datasets. The proposed fault-tolerance enhancement techniques of the CRAFT architecture are evaluated for various state-of-the-art DNNs using popular datasets. The DNN models used in the experiments include \emph{ResNet, VGG, MobileNet} and \emph{Inception}. The two datasets considered are \emph{CIFAR-10} and \emph{ImageNet}. The \emph{CIFAR-10} dataset contains 60,000 RGB images, of which, 50,000 images are for training and 10,000 images are for testing. Training set images are preprocessed by padding 4 pixels along the height and width and randomly cropping to a 32$\times$32 patch. Furthermore, random horizontal flip operation is also performed on the training set images. DNN models evaluated on \emph{CIFAR-10} dataset (i.e., ResNet-18, VGG-19 and MobileNet-V2) are trained using stochastic gradient descent with 0.9 momentum. The cross entropy loss function is used as the objective function to classify ten classes of the input images \cite{he2016deep}. The learning rate is kept constant at 0.1 during the training process.

Beside the \emph{CIFAR-10} dataset, the proposed method is also evaluated on the \emph{ImageNet} dataset, which consists of 1.3M images in the training set and 50,000 images in the test set. Training images are first randomly cropped to a 224$\times$224 patch and then followed by a random horizontal flip operation. Both training set and test set images are normalized using channel wise normalization for zero mean and unit standard deviation. During the training process, the network is learned by using the cross entropy loss as the objective function and minimized by the stochastic gradient descent algorithms with a momentum of 0.9. The learning rate is also kept at 0.1 during training on the \emph{ImageNet} dataset. After being trained, the network parameters are quantized to a lower precision (8-bit quantization) from 32-bit floating point representation. The baseline classification error of the evaluated models are shown in Table. \ref{table:1}. 

\begin{figure*}[t]
    \centering
    \includegraphics[]{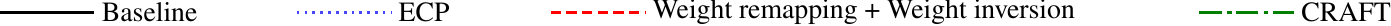}
\end{figure*}

\begin{figure*}[t]
	\centering
	\subfloat[]{\includegraphics{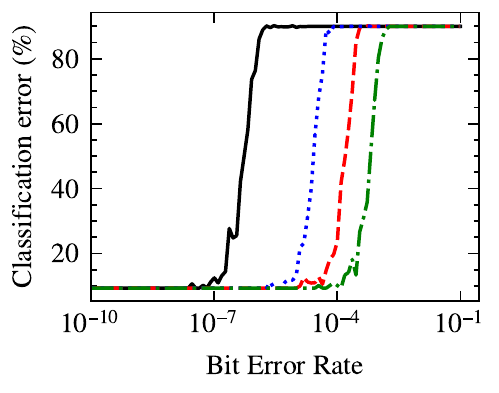}\label{fig:cifar10-resnet}}
	\qquad
	\subfloat[]{\includegraphics{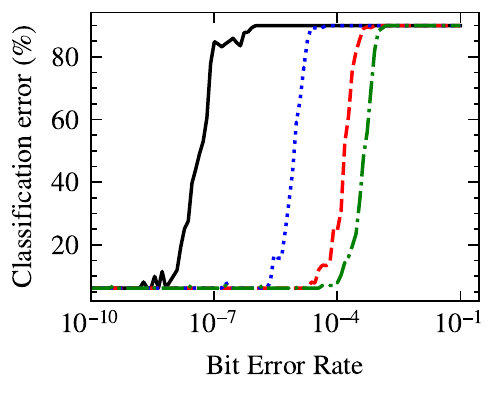}\label{fig:cifar10-vgg}}
	\qquad
	\subfloat[]{\includegraphics{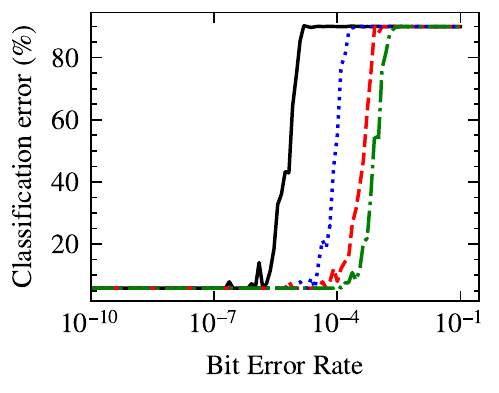}\label{fig:cifar10-mobilenet}}
	\caption{Comparison of different fault-tolerance techniques for various DNNs using CIFAR-10 dataset and 32-bit floating-point parameters   (a) ResNet-18  (b)  VGG-19  (c)  MobileNet-V2}
	\label{fig:cifar10-result}
\end{figure*}

\begin{figure*}[t]
	\centering
	\subfloat[\label{fig:resnet-imagenet}]{\includegraphics{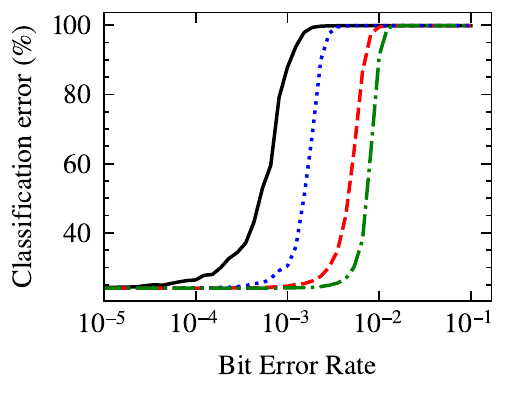}}
	\qquad
	\subfloat[\label{fig:vgg-imagenet}]{\includegraphics{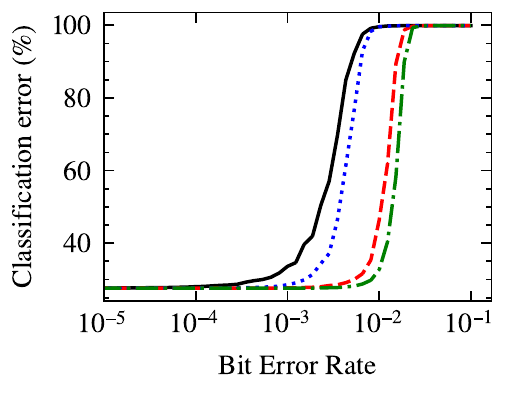}}
	\qquad
	\subfloat[\label{fig:inception-imagenet}]{\includegraphics{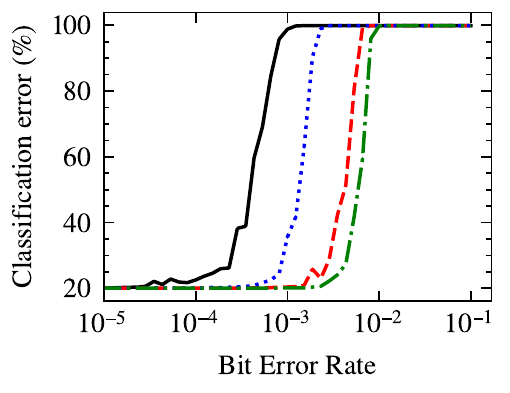}}
	\caption{Comparison of different fault-tolerance techniques for various DNNs using Imagenet dataset and 8-bit quantized parameters    (a)  ResNet-50  (b)  VGG-19  (c)  Inception-V4} 
	\label{fig:imagenet-result}
\end{figure*}

Quantization is a widespread technique, considered in many practical implementations, to reduce the computational overhead in DNNs. The network parameters are quantized using fewer bits resulting in a reduced precision. To evaluate the proposed method on quantized networks, we implement an 8-bit quantization scheme on DNNs that are evaluated on \emph{ImageNet} dataset. For quantized DNN models, a simple yet effective quantization scheme introduced in \cite{krishnamoorthi2018quantizing} is considered. Using this scheme, we implement 8-bit, weight only post-training quantization with asymmetric and per-layer mode.

\subsection{Stuck-at Faults Simulation Framework} 
Stuck-at faults in emerging memory-based DNNs are simulated using a customized fault-injection framework built upon the validated Ares framework introduced in \cite{reagen2018ares}. The experiments mainly consist of two stages. The first stage constructs various DNNs for experimentation using different weight precision and obtains possible parameter remappings/encodings based on the proposed methods presented in Sec. \ref{sec:proposed-techniques}.

In the second stage, the Ares framework is applied during DNN inference with the consideration of stuck-at faults in emerging NVMs. The stuck-at faults are assumed to be uniformly distributed in the network parameters, as mentioned in Sec. \ref{subsubsec:stuck-at-fault-in-nvms}. The proposed techniques are applied by considering the remapping/encoding configuration that leads to minimum net deviation in stage 1. For each bit error rate, the network classification error is averaged over 100 fault-injection simulations.

\subsection{Comparison with the Error Correcting Techniques}
The effectiveness of the proposed techniques is compared with the relevant existing Error Correcting Coding (ECC) solutions for the same purpose. As discussed earlier, the proposed techniques are flexible in terms of trade-off between the storage overhead and degree of robustness against stuck-at faults (more auxiliary bit can be used to enhance the fault-tolerance capability of the system at the cost of increased storage overhead). Therefore, we compare the proposed method with existing ECC techniques with similar storage overhead.

\begin{table*}[t]
	\centering
	\caption{Summary of Robustness Enhancement by different fault-tolerance techniques for various DNNs and datasets}
	\begin{tabular}{cccccccc} 
		\toprule
	    \multirow{2}{5em}{Dataset} & \multirow{2}{4em}{Model} & \multicolumn{4}{c}{Robustness Improvement } \\ 
                                   && Baseline & ECP & Weight Remapping + Weight Inversion\cite{nguyen_imran2021} & CRAFT \\
		\midrule
        \multirow{3}{5em}{CIFAR-10} & ResNet-18  & 1x & 81x & 351x & \textbf{1233x} \\ 
                                   & VGG-19 & 1x & 284x & 3511x & \textbf{12320x} \\
                                   & MobileNet-V2 & 1x & 23x & 53x & \textbf{231x} \\
		\midrule
        \multirow{3}{4em}{\centering ImageNet} & ResNet-50  & 1x & 4x & 15x & \textbf{29x} \\
                                               & VGG-19     & 1x & 2x & 8x & \textbf{12x}\\
                                               & Inception-V4  & 1x & 5x & 10x & \textbf{23x} \\
		\bottomrule
	\end{tabular}
	\label{table:2}
\end{table*}

In conventional memory system, ECC code such as (72, 64) Hamming code is often used for addressing soft errors. A typical (72, 64) Hamming code normally incurs more than 10\% of storage overhead, which is 10$\times$ larger than the combined overhead of all of the proposed techniques of CRAFT. Moreover, ECC schemes for addressing soft errors in conventional memory system are not as effective in emerging NVMs, as shown and discussed in \cite{10.1145/1815961.1815980}. Instead, Error Correcting Pointers (ECP) are often preferred for tolerating hard errors in emerging NVMs. For a fair comparison, we benchmark the proposed methods with the ECP variant that incurs a comparable storage overhead. For $d$-bit data, $ECP_n$ is able to correct up to $n$ bits with the storage overhead of $\frac{1 + n + n.\ceil{\log_{2} d}}{d}$. For our experiments, we consider $ECP_{1}$ which can correct a single bit error in 512-bit data block at the cost of $2.15\%$ storage overhead, which is still higher than that of CRAFT.

As discussed in Sec. \ref{sec:proposed-techniques}, for 32-bit data (one full-precision weight or four quantized weights), the proposed techniques together require approximately 1.17\% storage overhead. This amount is 2$\times$ less than the evaluated ECP. Despite having a minimal storage overhead, CRAFT is found to be more effective as compared to ECP, regardless of the DNN model type or dataset size. The detailed results are presented in the next section.

\subsection{Results and Discussion}\label{sec:results}

The evaluation results of the proposed methods for CIFAR-10 and ImageNet are shown in Fig. \ref{fig:cifar10-result} and Fig. \ref{fig:imagenet-result}, respectively. The solid black line shows the baseline models (considering stuck-at faults) without any error-correction scheme. The dotted blue line indicates the models which incorporate ECP to mitigate stuck-at faults. The red dashed line illustrates models that use \textit{Intra-Block Addressing Remapping} and \textit{Weight Inversion} technique for stuck-at fault-tolerance. The green dash-dotted line shows results for CRAFT architecture which includes \textit{Weight Remapping} + \textit{Weight Inversion} + \textit{Criticality-Aware Bits Switching}. As seen from the results, CRAFT improves the robustness of the baseline models significantly and outperforms the ECP method by a significant margin.

Specifically, Fig. \subref{fig:cifar10-resnet} shows the evaluation results of different fault-tolerance methods for ResNet-18 using CIFAR-10 dataset. The baseline and ECP classification error starts to increase exponentially at the bit error rate (BER) of $1.5\times10^{-7}$ and $5\times10^{-6}$, respectively. On the other hand, the classification error can be maintained below 10\% at around $5\times10^{-5}$ BER when Weight Remapping and Weight Inversion are applied and at $2\times10^{-4}$ when used together with Criticality-Aware Bits Swithcing in CRAFT. In other words, CRAFT can increase the fault-tolerance by up to more than 1200$\times$ compared to the baseline and 3$\times$ compared to only using Weight Remapping + Inversion methods. A similar trend can also be observed in other networks using the CIFAR-10 dataset such as VGG-19 and MobileNet-V2. For example, in VGG-19, while ECP can only improve the robustness of the model up to 284$\times$, CRAFT can increase the robustness to 12,320$\times$ compared with the baseline model, which is orders of magnitude higher than ECP. The efficiency of CRAFT compared to the baseline model when using MobileNet-V2 is found to be 231$\times$, which is 10$\times$ higher than ECP and 4$\times$ better than previously proposed methods \cite{nguyen_imran2021}.

To confirm the general applicability of the proposed techniques, we also perform different experiments on larger DNNs (ResNet-50, Inception-V4, etc\ldots) with larger dataset (i.e., ImageNet) (Fig. \ref{fig:imagenet-result}). As discussed in Sec. \ref{subsubsec:bit-pos-criticality}, the 8-bit quantized networks show a significant improvement in robustness compared to full-precision networks in Fig. \ref{fig:cifar10-result} due to the smaller dynamic range representation. Regardless of such property, the proposed techniques still ensure a significant increase in robustness against stuck-at faults, as shown in Fig. \ref{fig:imagenet-result}. For example, for ResNet-50 using ImageNet (Fig. \subref{fig:resnet-imagenet}), the network accuracy starts to drop at around $2\times10^{-4}$ BER while CRAFT can maintain the accuracy up to $5\times10^{-3}$ BER, indicating 29$\times$ more robustness than the baseline model. This trend is found to be consistent in other DNNs evaluated on ImageNet dataset. Specifically, CRAFT can enhance the robustness of VGG-19 and Inception-V4 up to 12$\times$ and 23$\times$, respectively. A summary of the robustness improvement over the baseline configurations using different fault-tolerance techniques is given in Table. \ref{table:2}. The improvement in robustness is obtained by measuring the BER at which the technique causes the classification error to increase by more than 5\%. As shown in the table, CRAFT outperforms the existing techniques by orders of magnitude (up to $10^{4}$ times over the baseline). While achieving a significant improvement in robustness against SAFs, as mentioned in Sec. \ref{subsec:implementation-overhead}, CRAFT incurs a minimum amount of storage overhead ($\approx$1.17\%), which makes CRAFT highly practical when implementing in resource-constrained edge devices.

Stuck-at faults in emerging NVMs are expected to become more frequent as technology scaling to smaller nodes. The two distinguishing features of CRAFT are consideration for the criticality of errors and flexibility to address more errors by employing a fine-grained remapping using smaller block size. This approach makes CRAFT a robust and scalable method which can tackle future trends of hard errors in NVMs. Finally, the proposed techniques of CRAFT are orthogonal to the existing error correcting techniques, therefore, further enhancement in the fault-tolerance can be achieved by implementing CRAFT along with the conventional techniques. 

\section{Conclusion}\label{sec:conclusion}
Hard errors such as stuck-at faults can severely impact the accuracy of Deep Neural Networks based systems which use emerging Non-Volatile Memories. This paper introduces a set of robust techniques, collectively named CRAFT, to enhance the error-tolerability of DNNs against stuck-at faults. The proposed techniques are simple and light-weight yet effective to tackle the problem of stuck-at faults in DNNs-based system. Working in a hierarchical manner, the proposed CRAFT architecture remaps the weights, encodes them using a simple inversion method and switches the bits within the weights based on their criticality, thus minimizing the impact of stuck-at faults on neural network's accuracy. The evaluation results show that CRAFT is able to enhance the robustness of the system by orders of magnitude. Specifically, for DNNs evaluated on CIFAR-10, CRAFT can increase the robustness by up to $10^{4}$ times compared to the baseline model. For DNNs using ImageNet, CRAFT enhances the robustness of the model by up to 29 times. Being orthogonal, the proposed techniques of CRAFT can be easily incorporated with other existing methods to further increase the fault-tolerance of DNNs. 

\bibliographystyle{IEEEtran}
\bibliography{references}

\begin{IEEEbiography}[{\includegraphics[width=1in, height=1.25in,clip,keepaspectratio]{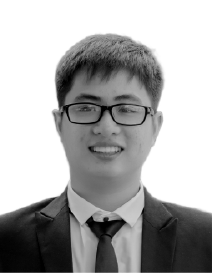}}]{Thai-Hoang Nguyen}
(Student Member, IEEE) received the BS degree in electrical engineering from Danang University of
Science and Technology, the University of Danang, Vietnam in 2019. He joined Sungkyunkwan
University, Suwon, Korea, in 2019 where he is currently working toward his PhD degree in electrical
and computer engineering. He is the recipient of STEM scholarship for international graduate student
by Sungkyunkwan University, from 2019. His current research interests include designing robust
architecture for emerging memory technologies, processing-in-memory and computationally efficient implementations of deep learning algorithms.
\end{IEEEbiography}

\begin{IEEEbiography}[{\includegraphics[width=1in, height=1.25in,clip,keepaspectratio]{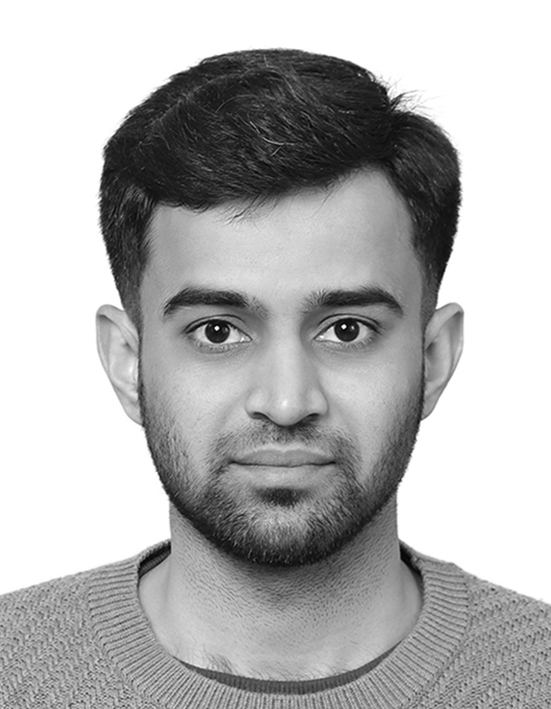}}]{Muhammad Imran} 
received the BS degree in electrical engineering from the University of Engineering and Technology,
Lahore, Pakistan, in 2012 and the PhD degree in electronic and electrical engineering from
Sungkyunkwan University, Suwon, South Korea, in 2020. He is currently an assistant professor at the
National University of Sciences and Technology, Islamabad, Pakistan. He is recipient of the Superior Research Award at Sungkyunkwan University and Scholarship by the Higher Education Commission of Pakistan for his MS and PhD studies. His current research interests include vector processor design and reliable computing architectures for deep learning.
\end{IEEEbiography}

\begin{IEEEbiography}[{\includegraphics[width=1in, height=1.25in,clip,keepaspectratio]{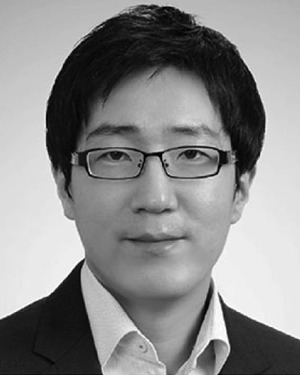}}]{Jaehyuk Choi}
Jaehyuk Choi (Associate Member, IEEE) received the B.S. degree in electrical engineering from Yonsei University, Seoul, South Korea, in 2004, the M.S. degree in electrical engineering and computer science from the Korea Advanced Institute of Science and Technology (KAIST), Daejeon, South Korea, in 2006, the M.S. degree in electrical and computer engineering from the University of Minnesota, Minneapolis, MN, USA, in 2008, and the Ph.D. degree from the University of Michigan, Ann Arbor, MI, USA, in 2013.
From 2013 to 2015, he was a Research Staff Member with the Samsung Advanced Institute of Technology (SAIT), Samsung Electronics, Suwon, South Korea, where he engaged in the development of depth sensors and low-power image sensors. In 2015, he joined the Department of Semiconductor Systems Engineering, Sungkyunkwan University, Suwon, as an Assistant Professor. His research interests include low-power circuits, CMOS sensors, and mixed-signal integrated circuits.
\end{IEEEbiography}

\begin{IEEEbiography}[{\includegraphics[width=1in, height=1.25in,clip,keepaspectratio]{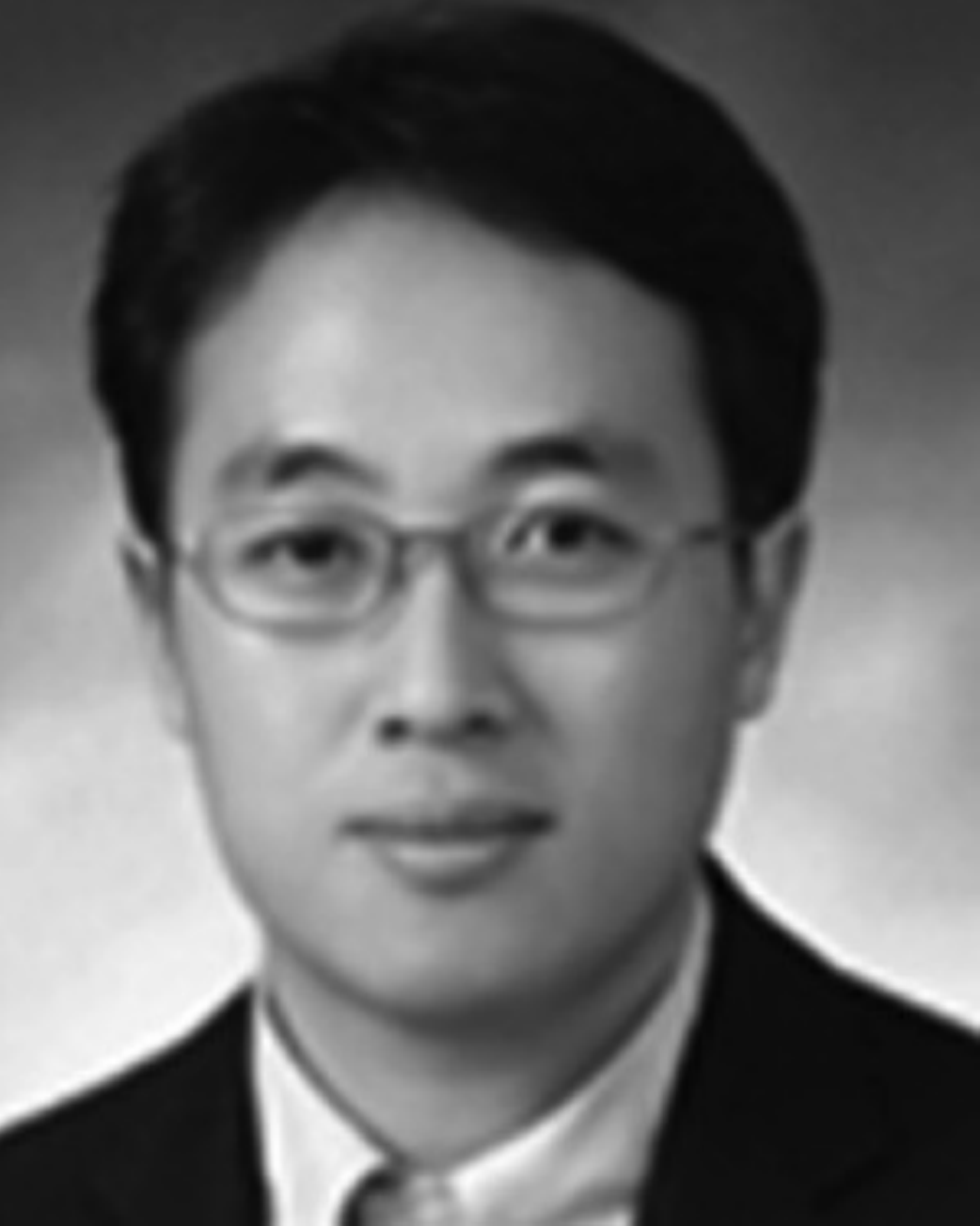}}]{Joon-Sung Yang}
Joon-Sung Yang (Senior Member, IEEE) received the B.S. degree from Yonsei University, Seoul, South Korea, in 2003, and the M.S. and Ph.D. degrees from The University of Texas at Austin, Austin, TX, USA, in 2007 and 2009, respectively, all in electrical and computer engineering. After graduation, he worked at Intel Corporation, Austin, for four years. He was with Sungkyunwan University. He is currently an Associate Professor with Yonsei University. His research interests include memory architectures and efficient deep learning architecture development. He was a recipient of the Korea Science and Engineering Foundation (KOSEF) Scholarship, in 2005. He received the Best Paper Award from the 2008 IEEE International Symposium on Defect and Fault Tolerance in VLSI systems and the 2016 IEEE International SoC Design Conference. He was nominated for the Best Paper Award from the 2013 IEEE VLSI Test Symposium.
\end{IEEEbiography}

\end{document}